\documentstyle[psfig,epsf]{mn}

\newcommand{\SS}{\scriptscriptstyle}
\newcommand{\Rl}{R_{\SS L1}}

\def\hel2{He\,{\sc ii}\,$\lambda$4686}
\def\hb{H$\beta$}
\def\hg{H$\gamma$}
\def\hu{HU~Aqr}
\def\uz{UZ~For}
\def\kmps{km~s$^{-1}$}

\begin{document}

\title{Accretion Stream Mapping of HU Aquarii}

\author[Sonja~Vrielmann and Axel~D.~Schwope]
{Sonja~Vrielmann$^1$\thanks{Send offprint requests to:
sonja@pinguin.ast.uct.ac.za},
Axel D.~Schwope$^2$\\
$^1$Department of Astronomy, University of Cape Town, Private Bag,
Rondebosch, 7700, South Africa\\
$^2$Astrophysikalisches Institut Potsdam, An der Sternwarte 16, 
14482 Potsdam, Germany}

\maketitle

\begin{abstract}

We present a new mapping algorithm, the {\em Accretion Stream Mapping
(ASM)} which uses the full phase-coverage of a light curve to derive
spatially resolved intensity distributions along the accretion stream
in magnetic cataclysmic variables of AM Herculis type (polars). The
surface of the accretion stream is approximated as a duodecadon shaped
tube. After successfully testing this method on artificial data we
applied it to emission line light curves of H$\beta$, H$\gamma$ and
\hel2\ of the bright eclipsing polar HU~Aqr.  We find Hydrogen and
Helium line emission bright in the threading region of the stream
where the stream couples onto magnetic field lines. It is particularly
interesting, that the stream is bright on the irradiated side,
facing the white dwarf, which highlights the interplay of collisional 
and radiative excitation/ionization.

\end{abstract}


\section{Introduction}
The strong magnetic field in cataclysmic binaries of AM Herculis type,
so called polars, keeps both stars in synchronous rotation (with a few
exceptions having slightly asynchronously rotating white dwarfs),
leads to intense cyclotron radiation from the small accretion spots
at the magnetic poles of the white dwarf
and prevents the formation of an accretion disc. Instead, stellar
matter is transferred via an accretion stream from the late-type
main sequence star (the secondary) to the white dwarf.

In the basic picture the stream leaves the secondary via Roche-lobe
overflow at the inner Lagrangian point $L_1$ to initially follow a
ballistic trajectory down to some radius where the magnetic pressure
overcomes the ram pressure (Liebert \& Stockman 1985).  In this
interaction region the growth of plasma instabilities of
Rayleigh-Taylor and Kelvin-Helmholtz type leads to a break-up of the
stream resulting in a fine rain of small droplets which follow
magnetic field lines and are finally accreted onto the white dwarf
(Hameury et al.~1986).

This scenario was observationally confirmed in its basic features by
Doppler tomography of bright emission lines revealing a ballistic
stream (Schwope, Mantel \& Horne 1997; SMH97 henceforth) and by the
occurence of X-ray absorption dips originating in the magnetic part of
the stream (e.g.~Watson et al.~1989).  X-ray emission from the stream
due to dissipative heating in the interaction region was predicted to
occur (Liebert \& Stockman 1985), but never observed.  Doppler
tomography also revealed the existence of an accretion curtain raising
from footpoints everywhere along the ballistic accretion stream.

Observationally, the question of locating the stream (or streams), the
accretion curtain and the brightness distribution along these
structures was pushed forward by mapping experiments of emission line
and continuum radiation in two eclipsing systems, HU~Aqr and UZ~For
(Hakala 1995, Harrop-Allin et al.~1998, 1999a, 1999b, Heerlein et
al.~1999, Kube et al.~2000, Sohl \& Wynn 1999, Vrielmann \& Schwope
1999).  Hakala and Harrop-Allin et al.\ were using broad-band UBVR
eclipse light curves of \hu\ in order to derive the brightness
distribution along the stream. They tried different stream geometries,
a single stream bound to the orbital plane or a stream with a
ballistic part and accretion down to one or two poles.  The major
drawback using this type of input data are the contributions of
various radiation components (i.e.\ the stream, the white dwarf and
the accretion spot with photospheric and cyclotron emission) with
different and partly unknown angular characteristics. Their advantage
is the high time resolution and good signal-to-noise ratio of the
broad-band data.

Kube et al.\ were using a similar geometry and optimization method as
the former authors, for the case of \uz. They mapped the brightness
distribution of the C{\sc iv}\,$\lambda$1550 emission line extracted
from high-time low-spectral resolution observations with the
HST/FOS around eclipse phase.
This approach has the advantage of eliminating the angle-dependent
continuum emission from the accretion spot.

%***********************************************************************
\begin{figure*}
\begin{minipage}{5cm}
\epsfxsize=5cm
\psfig{file=streamgeok.epsi,width=5cm}
\end{minipage}
\hspace{1cm}
\begin{minipage}{4cm}
\epsfxsize=4cm
\epsfbox{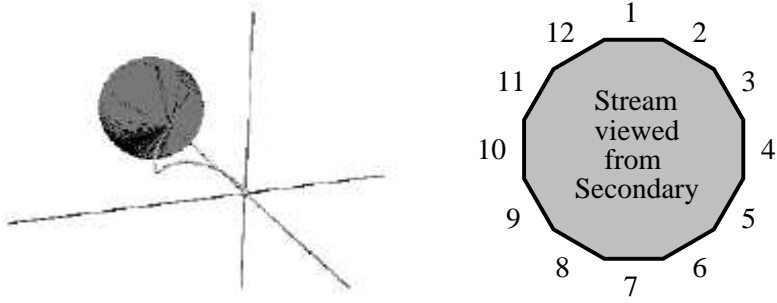}
\end{minipage}
\caption{{\em Left:} The geometry of the system.
Shown are the secondary star in the back, the accretion stream and the
white dwarf in a perspective view at binary phase $\simeq 0.55$. The 
coordinate axis is centred in the white dwarf. {\em
Right:} The duodecadon shaped stream 
crosssection as viewed from the inner Lagrangian point L$_1$ on the
secondary.
\label{f:geo}}
\end{figure*}
%***********************************************************************

Heerlein et al.~and Sohl \& Wynn modelled the full, complex emission
line profile of \hel2\ in \hu\ obtained with full phase coverage and
high time and spectral resolution when the system was in a high
accretion state.  The resulting maps are constrained by the velocity
distribution observed at all phases rather than by the brightness
variations during eclipses. Hence, these methods give complementary
information to the eclipse mapping methods.

Here, we present a new approach to derive the brightness distribution
along the accretion stream in \hu\ taking the three-dimensionality of
the stream into account. 
We analyse emission line light curves of
Balmer \hb, \hg, and \hel2, obtained with full phase-resolution and
subtracted reprocessed radiation from the secondary star (SMH97).
Schwope et al.\ have shown that the stream emits differently on
its irradiated (inner) side than on its non-irradiated (outer)
side. We, therefore, map in our model the observed brightness
distribution on an accretion stream approximated as an opaque tube.
Some early results were presented in Vrielmann \& Schwope (1999).  

The structure of the paper is the following: in Sect.~\ref{s:data} we
describe the data used for the mapping method; Sections~\ref{s:method}
and \ref{s:sys_para} outline our model and assumptions; the
application to artificial data is described in Sect.~\ref{s:app_art};
and to observed data in Sect.~\ref{s:app_hu}. The paper is closed by a
discussion section~\ref{s:disc}, where our results are compared to
those of other groups.

\section{The data}\label{s:data}
Spectroscopy of \hu\ with high time and spectral resolution was
performed in the night August 17/18, 1993, with the double-beam
spectrograph TWIN at the 3.5m telescope on Calar Alto,
Spain. Simultaneous high-speed UBVRI photometry was performed at the
2.2m telescope, only 300\,m away. The photometry allowed us to reduce
trailing errors and seeing variability in the spectroscopic data. The
individual reduction steps are described in SMH97, where a detailed
analysis of the emission lines, in particular \hel2, is presented.

For our present analysis we use the light curves in integrated light
of H$\beta$, H$\gamma$, and \hel2. Since in our present study we are
interested only in emission originating from the accretion stream, we
corrected the integral light curve for the emission orginating on the
irradiated surface of the secondary. For that purpose we used the
Gaussian deconvolution of the \hel2\ line (SMH97, Fig.~7), smoothed
the observed asymmetric light curve of the narrow emission line from
the secondary star and subtracted it from the integrated light
curve. We used the same correction function, appropriately scaled, for
all three lines concerned.  The remaining emission is regarded to
originate solely from the accretion stream; the resulting light curves
are shown in Fig.~\ref{f:results}.  We will discuss below
(Sect.~\ref{s:app_hu}) the effects of uncertainties of the correction
function.

\section{Accretion Stream Mapping, ASM}
\subsection{The method}\label{s:method}

The {\em Accretion Stream Mapping} method, henceforth called ASM, is
based on classical {\em Eclipse Mapping} of accretion discs in
eclipsing cataclysmic variables (Horne 1985). This classical algorithm
is used to reconstruct intensity distributions in the accretion disc
by fitting the eclipse profile and applying a maximum entropy
algorithm in order to retrieve a unique solution for this otherwise
ill-conditioned problem.

Two major differences which are implemented in ASM with respect to the
classical {\em Eclipse Mapping} are:
(a) the disc geometry is replaced by a stream geometry;
(b) since the accretion stream eclipses {\it itself}, we use the full
	orbital light curve, instead of a short data interval centred
	on eclipse.
In principle, we are not confined to eclipsing systems (i.e.\ systems
which show an eclipse of the white dwarf by the secondary; see also
Section~\ref{s:test2}), however, they provide us with valuable
information in form of ingress and egress profiles in the light
curve. Therefore, we use data of the eclipsing polar HU~Aqr.

As a consistency check for our solution we use B-band photometry with
high time resolution (0.5 sec) around eclipse for comparison. Our
approach assumes 
that emission from the stream is completely optically thick, the
observed brightness variation is then solely due to projection and
occultation of stream pixels and not to variable optical depth.

The stream is modelled as a tube with twelve sides (duodecadon shaped)
centred on a one-particle trajectory under the influence of a magnetic
field.  Leaving the inner Lagrangian point $L_1$, the motion is
initially purely ballistic and the trajectory bound to the orbital
plane.  Later the matter connects to the magnetic field of the white
dwarf. The azimuth of the threading point is determined by the centre
of the optical/X-ray absorption dip in the light curve of \hu, just
preceeding eclipse. This dip is caused by the magnetic part of the
stream close to the threading region crossing the line of sight to the
white dwarf. From the threading region onwards, the stream is assumed
to follow a dipolar field line.

The geometry of the non-transparent stream is shown in
Fig.~\ref{f:geo}. Pixel 1 is always facing ``upwards'', i.e.\ the
surface normal of pixel 1 is parallel (in the ballistic stream) 
or closest to the orbital axis (in the magnetic stream). 
The area normal of pixel 4 is pointing {\em outwards}, that of
pixel 10 {\em inwards} with respect to the white dwarf. At any given
time, maximally half the pixels can be seen by the observer. However,
the identity of the visible pixel change with orbital phase. This
justifies the use of the {\em full observed light curve}. In eclipsing
system like HU~Aqr, the pixels are additionally occulted by the
secondary during a short phase interval.

The length of the trajectory is divided into 110 parts of equal
length, hence our resulting maps will have $12 \times 110$ pixels.  For
simplicity, the eclipse of the rising part of the magnetic stream by the
falling part, a short-time event, and vice versa is not included.

The phasedependent (integrated) flux $F_{\nu,\varphi}$ at frequency
$\nu$ in the light curve is calculated from the map as:
\begin{equation}
F_{\nu,\varphi} = f(d) \sum_{j=1}^{M} (I_j \,\, v_{j,\varphi} \,\,
a_{j,\varphi})
\end{equation}
where $f(d) = 10^{26} \Rl^2 / d^2 \cos i$ is a scale factor depending
on the distance $d$ and the size of the system expressed as $\Rl$ in
cgs units, as well as including a conversion from intensity to flux
units (factor $10^{26}$) and the inclination angle $i$. $I_j$ is the
intensity at the pixel $j$ in the stream, $M$ the number of pixel,
$v_{j,\varphi}$ the occultation function, $a_{j,\varphi}$ the
projected area of the pixel and $\varphi$ denotes the orbital phase.

The data are fitted according to the maximum entropy method (MEM,
Skilling \& Bryan 1984) by minimizing the quality function Q (Horne
1985, Bryan \& Skilling 1984):
\begin{equation}
Q = \frac{1}{N}\sum_{\varphi=1}^N \left(\frac{f_\varphi -
F_\varphi}{\sigma_\varphi}\right)^2 - \alpha S
\end{equation}
where $f_\varphi$ is the observed flux, $p_\varphi$ the predicted
flux, $\sigma_\varphi$ the uncertainty of the observed flux and $N$
the number of data points in the light curve. $S$ is a quantity
describing the entropy of the intensity distribution along the stream
and calculated as
\begin{equation}
S = \sum_{j} {I_{j} - D_{j} - I_{j} \ln{\frac{P_{j}}{D_{j}}}}
\end{equation}
where $D_{j}$ is the so-called {\it default} image calculated as a
smeared out version of the actual image $I_j$ and $\alpha$ is the
Lagrange multiplier. In ASM the smearing of the {\it default} image is
performed in two dimensions, i.e.~around and along the stream tube.
The chosen way of smearing is the only sensible one. Imagine a spot on
the stream, then it can be apparently smeared out due to optical
depths effects around the stream and/or smeared along the stream due to
the motion of the material in the stream.

MEM uses the information of the gradients to find the minimum of
the quality function $Q$. The criterion to stop iterating is defined
by convergence, i.e.\ when the gradients corresponding to the $\chi^2$
and the entropy $S$ are antiparallel. In a rough picture, this ensures
us of being in the entropy-$\chi^2$-parameter space right inbetween
the maximum of entropy and the minimum of the $\chi^2$. A further
explanation of this convergence criterion can be found in Bryan \&
Skilling (1984) and Horne (1985).

MEM guarantees that the solution we retrieve is the simplest, but
still compatible with the observations. However, there is no guarantee
that we always find the global minimum and do not end up in a local
minimum with little chance to escape from it. To ensure a minimum of
prejudice and allow comparison of different maps, the inversion
process was always started with an absolute flat, homogeneous map. The
reconstructions of artificially produced maps (as done in
Section~\ref{s:app_art}) helps us to learn the behaviour of the
method.

\subsection{The system parameters}\label{s:sys_para}

The geometry of the stream and our aspect on it are determined by five
parameters: the mass ratio $Q = M_{\rm wd}/M_2$, the orbital
inclination $i$, the orientation of the magnetic axis (colatitude
$\delta_{\mu}$, azimuth $\chi_{\mu}$) and the azimuth $\chi_{\rm th}$
of the threading region.  All these parameters are reasonably well
determined from optical and X-ray photometry and spectroscopy. We have 
checked, that the
remaining uncertainties in these parameters do not affect our results.
All azimuthal angles are measured in a reference system centred on the 
white dwarf with respect to the line joining both stars. The inner 
Lagrangian point $L_1$ has $\chi = 0\degr$, the accretion stream is deflected
to negative azimuths.

Unpublished spectroscopy of photospheric Na-lines from the secondary star shows
that the mass ratio must be higher than $Q > 3.5$. We adopt $Q=4$
throughout our analysis. The optical eclipse width then constrains the
inclination to $i = 85.6\degr$.

The X-ray bright phase of \hu\ as observed with ROSAT in a high
accretion state in October/November 1993 is centred on phase
$\phi_{\rm ecl} = 0.875$. Assuming that the azimuth for the accretion
spot $\chi_{\rm spot}$ and the dipolar axis $\chi_{\mu}$ are the same,
this allows us to derive the azimuth of the magnetic axis to
$\chi_{\mu} \simeq \chi_{\rm spot} = -45\degr$.  Optical and X-ray light curves
show a pronounced pre-eclipse dip centred on $\phi_{\rm ecl} = 0.880$. 
The dip is caused by absorption of radiation from the white dwarf 
in the magnetic stream. Since HU~Aqr is a high-inclination system, 
the phase of the dip indicates the azimuth $\chi_{\rm th}$ of the 
threading region, $\chi_{\rm th} \simeq -43\degr \simeq \chi_{\mu}$.  
The magnetic co-latitude of the dipolar axis finally was
derived by SMH97 and Heerlein et al.~(1999) to be of the order $10\degr -
15\degr$.

It was mentioned already in the introduction, that an accretion
curtain is likely present in \hu. This was derived by SMH97 from the
observed shielding of the leading hemisphere of the secondary star
from X-ray irradiation.  The curtain is expected not only to absorb
continuum radiation, but also to emit line radiation. The adoption of
a stream instead of a curtain is therefore oversimplified, but
unavoidable due to the one-dimensionality of the input data. We tried
to map the observed brightness distribution also on a simple accretion
curtain with the result, that the intensity is not distributed over
the full curtain (as expected), but that only some curious pixel become
bright. Mapping on a full curtain is therefore more complicated and
requires to take e.g.\ velocity information into account.

\subsection{Application to artificial data}
\label{s:app_art}
In order to evaluate how well our new method can reconstruct the
stream emission, we tested it on artificial data.  For that purpose we
computed artificial intensity distributions, calculated the
corresponding light curves and reconstructed the initial intensity
distributions by ASM. From the deviations between the reconstructions
and the original intensity distributions we learn about the behaviour
of the algorithm, its power and limits.

\subsubsection{Tests for different spot locations}
\begin{figure*}
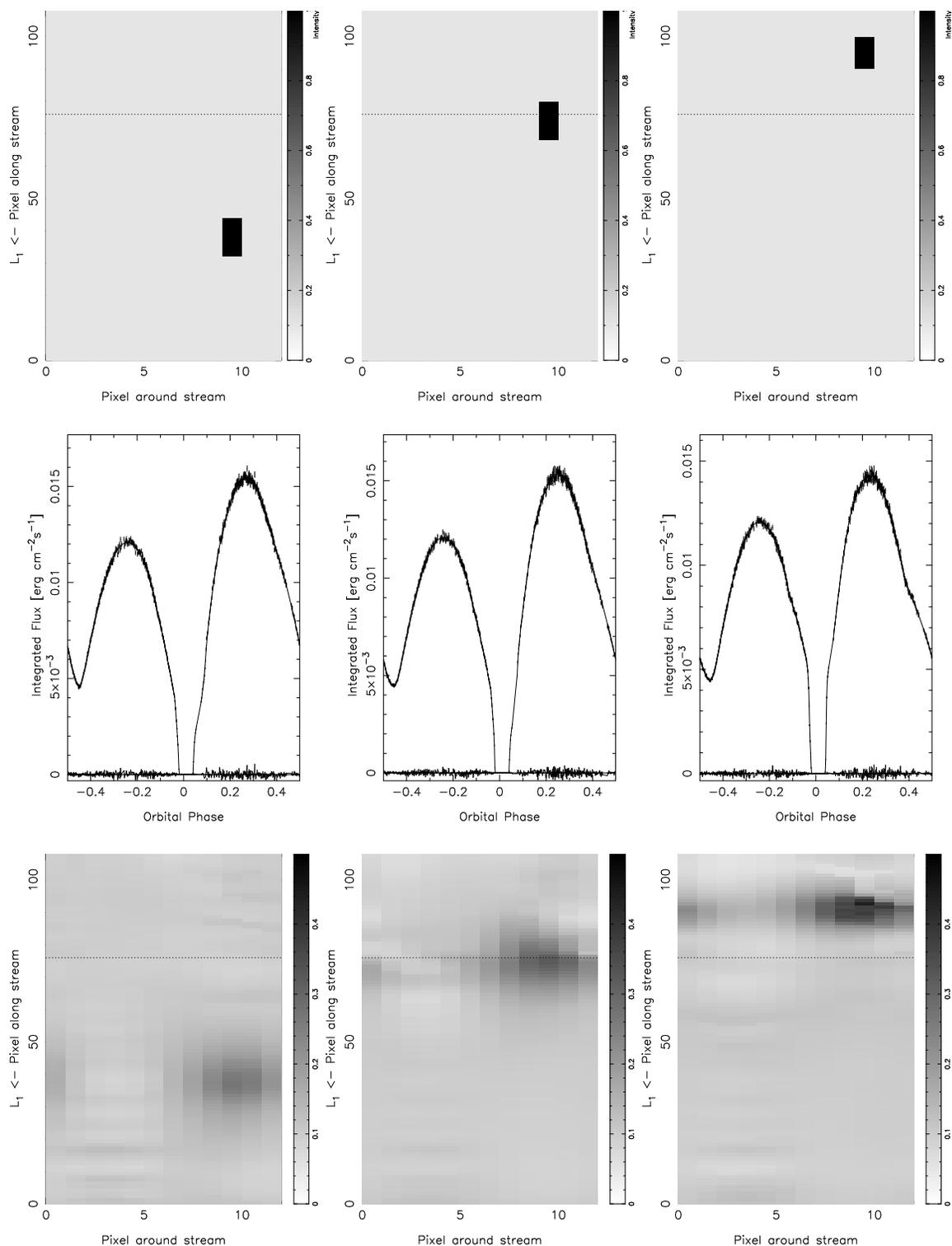

\begin{minipage}{5cm}
\psfig{file=stream10_67tv.epsi,width=5cm,%
bbllx=47pt,bblly=34pt,bburx=567pt,bbury=692pt,clip=}
\end{minipage}
\begin{minipage}{5cm}
\psfig{file=stream10_34tv.epsi,width=5cm,%
bbllx=47pt,bblly=34pt,bburx=567pt,bbury=692pt,clip=}
\end{minipage}
\begin{minipage}{5cm}
\psfig{file=stream10_12ntv.epsi,width=5cm,%
bbllx=47pt,bblly=34pt,bburx=567pt,bbury=692pt,clip=}
\end{minipage}

%light curves************************************************************
\vspace{0.5cm}
\begin{minipage}{4.4cm}
\psfig{file=stream10_67f.epsi,width=4.4cm}
\end{minipage}
\hspace{0.5cm}
\begin{minipage}{4.4cm}
\psfig{file=stream10_34f.epsi,width=4.4cm}
\end{minipage}
\hspace{0.5cm}
\begin{minipage}{4.4cm}
\psfig{file=stream10_12nf.epsi,width=4.4cm}
\end{minipage}

%result maps************************************************************
\vspace{0.5cm}
\begin{minipage}{5cm}
\psfig{file=stream10_67v.epsi,width=5cm,%
bbllx=47pt,bblly=34pt,bburx=567pt,bbury=692pt,clip=}
\end{minipage}
\begin{minipage}{5cm}
\psfig{file=stream10_34v.epsi,width=5cm,%
bbllx=47pt,bblly=34pt,bburx=567pt,bbury=692pt,clip=}
\end{minipage}
\begin{minipage}{5cm}
\psfig{file=stream10_12v.epsi,width=5cm,%
bbllx=47pt,bblly=34pt,bburx=567pt,bbury=692pt,clip=}
\end{minipage}
\caption{Test of ASM for different spot locations using artificial
data: Shown are the input maps (top), the original and reconstructed
light curves (middle) and the reconstructed maps (bottom). The maps
have the pixel-coordinate around the stream along the abscissa
(according to the convention shown in Fig.~\ref{f:geo}) and the
pixel-coordinate along the stream at the ordinate. The horizontal
dashed line corresponds to the threading region.  On input, a constant
intensity distribution with a bright spot located on the ballistic
stream (left), the coupling region (centre), and the magnetic stream
(right) has been assumed. Noise has been artificially created with a
signal to noise of 100 and added to the data before reconstruction.
\label{f:test}}
\end{figure*}

For these tests we used the same stream and viewing geometry as for
the real data of \hu\ as explained in the previous Section.  We
created three different artificial stream maps, each with a spot on
the inner side of the stream (pixel row 10), but different locations
along the stream. In the first case, the spot is located in the middle
of the ballistic part of the stream, in the second case at the
threading region and in the third case on the magnetic part of the
stream. Outside the spot, the intensity was set to a constant level of
10\% of the spot intensity.  For these artificial pixel patterns
(Fig.~\ref{f:test}, top row) we calculated the light curves with the
same phase resolution (and distribution) as in the observed data and
added artificial noise with a signal-to-noise (S/N) of 100
(Fig.~\ref{f:test}, middle row) plus a constant error, to avoid
vanishing errors at mid-eclipse. These synthetic data were then
analysed with ASM (Fig.~\ref{f:test}, bottom row).

%input maps************************************************************
%***********************************************************************

The synthetic light curves are determined by the folding products of
the phase-dependent visibility and the foreshortening functions of
each pixel.  These are then summed over all pixels for a given phase.
Consequently, the models with spots on the ballistic part of the
stream do not show any difference in the phase interval $\simeq 0.55 -
1.00$: at these phases the spot is invisible for the observer ($i
\simeq 85\degr$).  Differences exist between these two light curves at
egress phase, where the light curve of the model with the spot closer
to the $L_1$-point shows a more pronounced step-like behaviour, and
for the phase of the primary maximum. This maximum appears latest in
the binary cycle for the model with the spot closer to $L_1$ which
is the natural consequence of the curvature of the trajecory: the
observer has to move by the largest angle around the binaries axis
until the optimum projection angle for the region with the bright spot
is reached.

The third model with the spot on the magnetic part of the stream
differs at all phases from the first two models: its eclipse light
curve is much more symmetric than those of models '1' and '2'; the
primary maximum appears at earliest phase; the ingresses to the
primary and secondary mimina show each a bend not present in the other
two light curves; and the contrast between the primary and the
secondary maximum is reduced.

In the bottom row of Fig.~\ref{f:test} we show gray-scale images of
the reconstructed intensity distributions. The {\em location of the
spot} is in each case reconstructed very well, the spots are only
smeared out due to the MEM algorithm, leading to a different spot
profile and lower spot intensity.

We find some low amplitude structures near the secondary star on the
outer side of the stream (circumference pixel $1-7$, length pixel
$<50$).  These structures were probably introduced due to fitting
noise in the ingress phase, the only phase when this region is
visible. If we see such structures in reconstructions of real light
curves, they might be caused similarly and therefore have to be
treated as artefacts.

%light curves************************************************************
\begin{figure*}
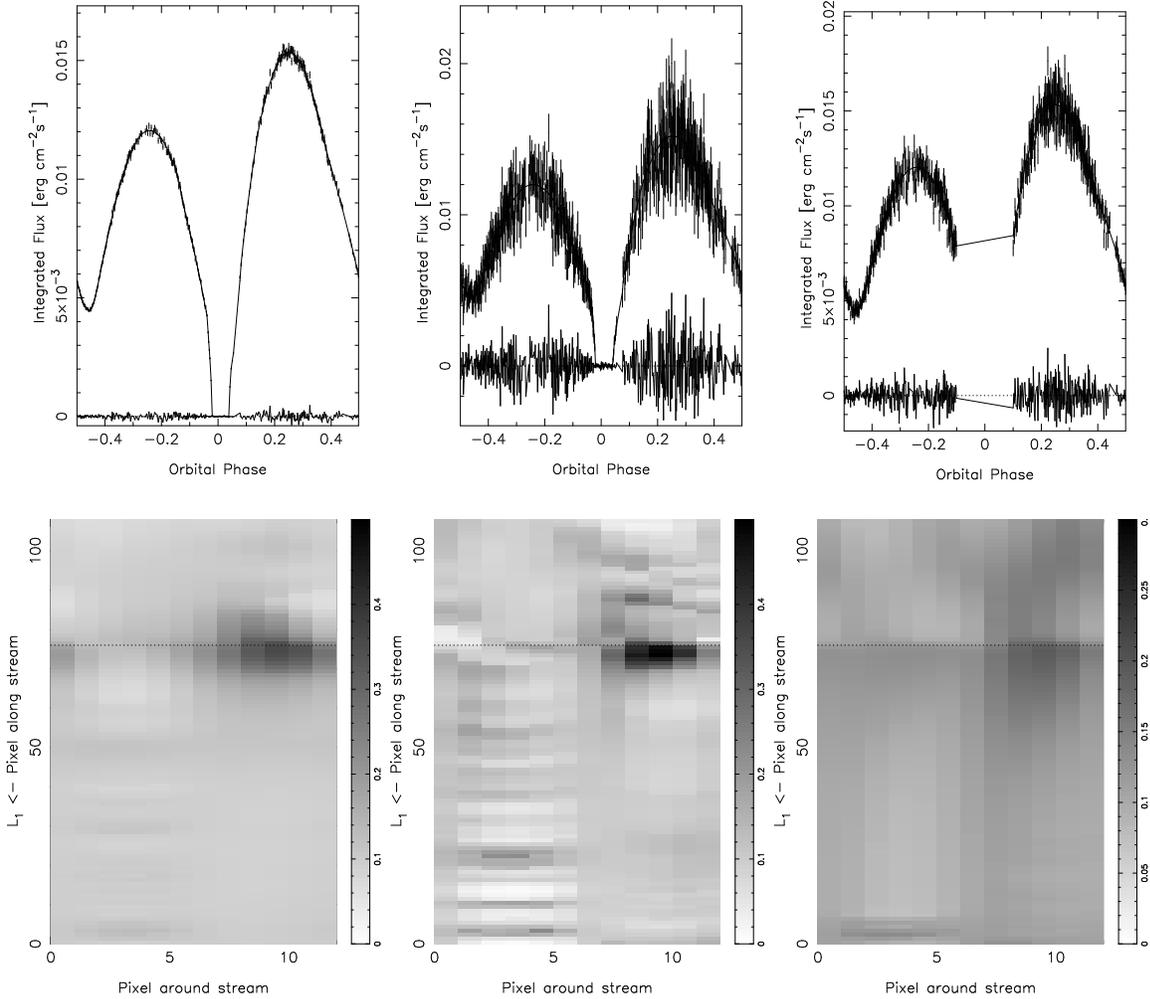

\begin{minipage}{4.4cm}
\psfig{file=test34_361f.epsi,width=4.4cm}
\end{minipage}
\hspace{0.5cm}
\begin{minipage}{4.4cm}
\psfig{file=tstn10f.epsi,width=4.4cm}
\end{minipage}
\hspace{0.5cm}
\begin{minipage}{4.4cm}
\psfig{file=tstnes1f.epsi,width=4.4cm}
\end{minipage}

%result maps************************************************************
\vspace{0.5cm}
\begin{minipage}{5cm}
\psfig{file=test34_361v.epsi,width=5cm,%
bbllx=47pt,bblly=34pt,bburx=567pt,bbury=692pt,clip=}
\end{minipage}
\begin{minipage}{5cm}
\psfig{file=tstn10v.epsi,width=5cm,%
bbllx=47pt,bblly=34pt,bburx=567pt,bbury=692pt,clip=}
\end{minipage}
\begin{minipage}{5cm}
\psfig{file=tstnes1v3.epsi,width=5cm,%
bbllx=47pt,bblly=34pt,bburx=567pt,bbury=692pt,clip=}
\end{minipage}
\caption{Test of ASM with lower quality data:
using a twice as low phase 
resolution (left), a ten times as low signal-to-noise ratio (S/N=10) (middle)
compared to Fig.~\ref{f:test} and a missing eclipse coverage (right;
the line in the light curve accross the eclipse simply connects the
flux points at phases $-0.1$ and $0.1$). The input map is identical to the
one in Fig.~\ref{f:test} (top middle) with a spot located on the
coupling region. Shown are the original and reconstructed light curves
(top) and the reconstructed maps (bottom). The maps have the
pixel-coordinate around the stream along the abscissa (according to
the convention shown in Fig.~\ref{f:geo}) and the pixel-coordinate
along the stream at the ordinate. The horizontal dashed line
corresponds to the threading region.
\label{f:test2}}
\end{figure*}
%***********************************************************************

\subsubsection{Tests with lower quality data}
\label{s:test2}

Additionally, we tested the method for lower phase resolution, a lower
signal-to-nose ratio and a missing eclipse coverage as shown in
Fig.~\ref{f:test2}.

For the first test with low quality data we set the phase resolution
simply to twice as low as that in the first test cases by using only
every second phase value. This especially affects the egress phase
which already has a low phase resolution.  Nevertheless, we can
recover the spot very well and in general is the reconstruction little
different from the one in Fig.~\ref{f:test} (middle bottom). The only
difference is an enhancement of artefacts, e.g.\ the low amplitude
structures near the secondary star (see above).  Even with a phase
resolution of only 20 data points during the full orbit could we
recover indication of the spot, however, severely smeared out around
the stream.

In the second of these tests we set the signal-to-noise ratio
artificially to a value of 10, i.e.~ten times lower as for the tests
in Fig.~\ref{f:test} and comparable to the signal-to-noise ratio in
the HU~Aqr data. The constant additional error was also set a factor
of 10 higher than in the other test cases. Still, we recover the spot
very well and the reconstruction differs only in terms of, though
strongly, enhanced artefacts near the secondary (as described above)
and on the magnetic part of the stream.

In the third of these test cases we used the same phase resolution and
phase coverage as in Fig.~\ref{f:test} (middle middle), but cut out
the fluxes between phases $-0.1$ and $0.1$. Additionally, we used here
a S/N of 20 instead of 100, as in most other test cases. Even if the
eclipse is not covered by observation, we can still see indication of
the spot at the threading region, however very weak and smeared out
along and around the stream. The artefacts seen in all other
test cases disappeared, indicating that the data points around the
eclipse are truely responsible for them. The lower signal-to-noise
ratio does not have any large effect on the reconstuction, the maps
for S/N = 100 and S/N = 20 look very similar. This test shows very
well, how the out-of-eclipse light curve is determined by the
emissivity of the stream. While the eclipse profile gives details and
sharper contrasts in the maps, the general behaviour of the outside
eclipse light curve provides us with information on the general
features in the stream emissivity: the difference in maxima at phases
$-0.25$ and 0.25 indicate which side of the stream is brighter and the
shape of the (brighter local) maximum (spanning about a third of the
orbit) indicates, where the spot is (cf.\ also the light curves in
Fig.~\ref{f:test}). This information is little influenced by flickering
or a low signal-to-noise ratio.

These tests show that our new method is very robust and can recover
the general features of the original intensity distribution even from
lower quality data. {\em This is due to the use of the full light
curve.} Of course, the quality of the reconstructions depends on using
the right geometry for the stream. This is shown in test
reconstructions in the following Section.

%light curves************************************************************
\begin{figure*}
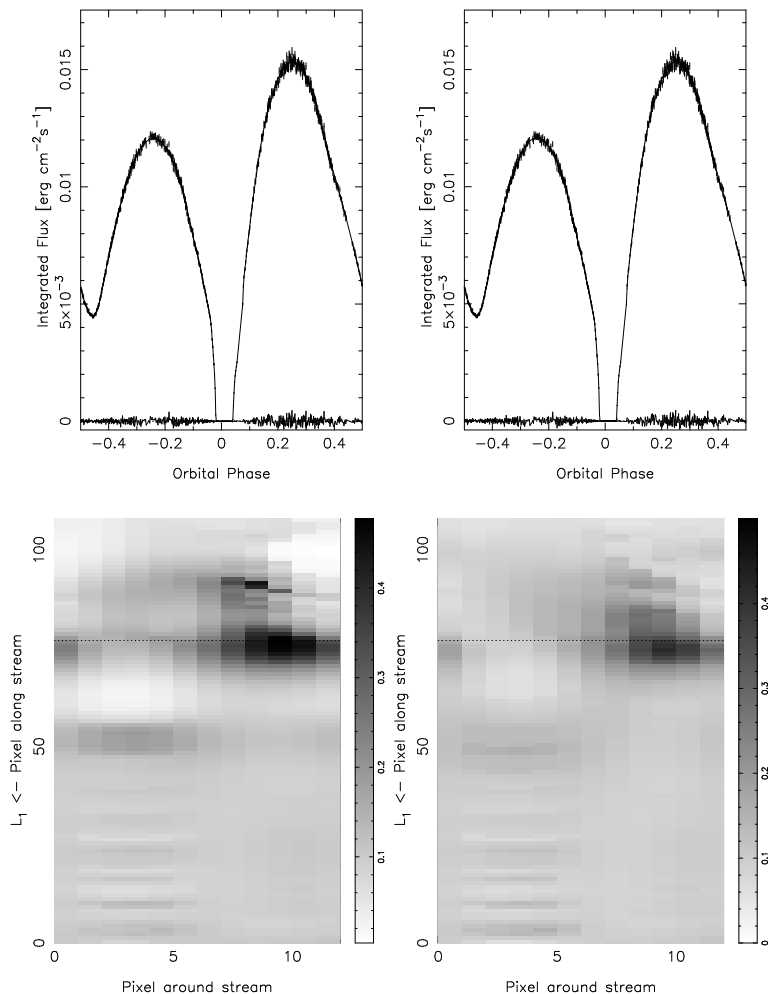

\begin{minipage}{4.4cm}
\psfig{file=tst3_1f.epsi,width=4.4cm}
\end{minipage}
\hspace{0.5cm}
\begin{minipage}{4.4cm}
\psfig{file=tst4_1f.epsi,width=4.4cm}
\end{minipage}

%result maps************************************************************
\vspace{0.5cm}
\begin{minipage}{5cm}
\psfig{file=tst3_1v.epsi,width=5cm,%
bbllx=47pt,bblly=34pt,bburx=567pt,bbury=692pt,clip=}
\end{minipage}
\begin{minipage}{5cm}
\psfig{file=tst4_1v.epsi,width=5cm,%
bbllx=47pt,bblly=34pt,bburx=567pt,bbury=692pt,clip=}
\end{minipage}
\caption{Test of ASM for the stream geometry: using a too small
colatitude (left) and a too small azimuth (right) of the
magnetic axis.  The input light curve is identical to the one in
Fig.~\ref{f:test} (middle middle) created from a map with a spot
located on the coupling region. Shown are the original and
reconstructed light curves (top) and the reconstructed maps
(bottom). The maps have the pixel-coordinate around the stream along
the abscissa (according to the convention shown in Fig.~\ref{f:geo})
and the pixel-coordinate along the stream at the ordinate. The
horizontal dashed line corresponds to the threading region.
\label{f:test3}}
\end{figure*}
%***********************************************************************

\subsubsection{Tests for the stream geometry}
\label{s:test3}

Furthermore, we tested the ASM method against choosing wrong
parameters for the stream geometry. For these tests we used the light
curve in Fig.~\ref{f:test} (middle, middle) and reconstructed the
intensity distribution assuming slightly wrong parameters for the mass
ratio, the azimuth of the threading region (determined by the magnetic
field strength), the colatitude and the azimuth of the dipolar axis.

In the test case where we chose an azimuth of the threading region
smaller by about $8^\circ$ (i.e.\ a magnetic field strength smaller by
about 15\%) the magnetic part of the stream also becomes shallower,
changing the geometry quite dramatically. For this test case we cannot
find any converged solution, not even for an as high $\chi^2$ as
10. The intermediate, unconverged map shows artefacts in form of
pixel-to-pixel variations. This means the azimuth of the threading
region is very well defined.

Similar is the behaviour of the reconstruction process when we choose
a false mass ratio. For the test case we used $q = Q^{-1} = 0.3$. It
changes the stream geometry in a similar way as in the previous test
case, however in the opposite direction. Again we could not find a
converged solution for even a $\chi^2$ of 10. The intermediate
unconverged maps show again strong artefacts, however, indication for
the spot at the right location.

If we change the colatitude of the magnetic axis, it mainly changes
the spatial amplitude of the magnetic part of the stream. A value
smaller by $5^\circ$ leads already to a 10\% increase in
deflection. The resulting map (Fig.~\ref{f:test3}) still shows the
spot, but also an additional bright spot at the {\em inner} side of
the magnetic part of the stream and a dark spot on the {\em upper}
side. While this alone would not allow us to dismiss the geometry, the
pixel-to-pixel variation around the second spot and enhanced artefacts
on the ballistic part of the stream indicate that we chose a false
geometry. The appearance of this second spot is understandable: the
reconstructed intensity distribution on the magnetic part of the
stream for the too strongly deflected stream path is asymmetric in an
attempt to reproduce the true intensity distribution of the shallower
stream path. Bright or dark spots around the spatial maximum of the
stream path should therefore be interpreted cautiously. They can be
introduced by wrong geometry parameters.

The azimuth of the magnetic axis is much less well defined than the
other geometry parameters. Only for a angle different by more than
$30^\circ$ we could see a significant change in the deflection in the
magnetic part of the stream (and a shift in the location of the
maximum of spatial deflection). The map with the reconstructed
intensity distribution as shown in Fig.~\ref{f:test3} therefore shows
not very extreme differences to the one in Fig.~\ref{f:test} (middle
bottom), but also indication for a second spot on the magnetic part of
the stream and enhanced artefacts.

These tests show that the geometry of the system is relatively well
defined. In case we choose false geometry parameters (especially false
values for the magntic field or the mass ratio) the reconstructions
show artefacts and pixel-to-pixel variations. Such structures will
give us some indication for the goodness of the geometry parameters
from the final reconstructions. If on the other hand a geometrical
parameter is less well defined (as the azimuth of the magnetic axis),
it does not influence the map to a great extend and the general
features are reconstructed very well.

%light curve fits************************************************************
\begin{figure*}
\begin{minipage}{5.0cm}
\psfig{file=hb_sh_7f.ps,width=5.0cm}
\end{minipage}
\hfill
\begin{minipage}{5.0cm}
\psfig{file=hg_sh_7f.ps,width=5.0cm}
\end{minipage}
\hfill
\begin{minipage}{5.0cm}
\psfig{file=he_sh_4f.ps,width=5.0cm}
\end{minipage}
\hfill

%result maps************************************************************
\vspace{0.5cm}
\begin{minipage}{5.0cm}
\psfig{file=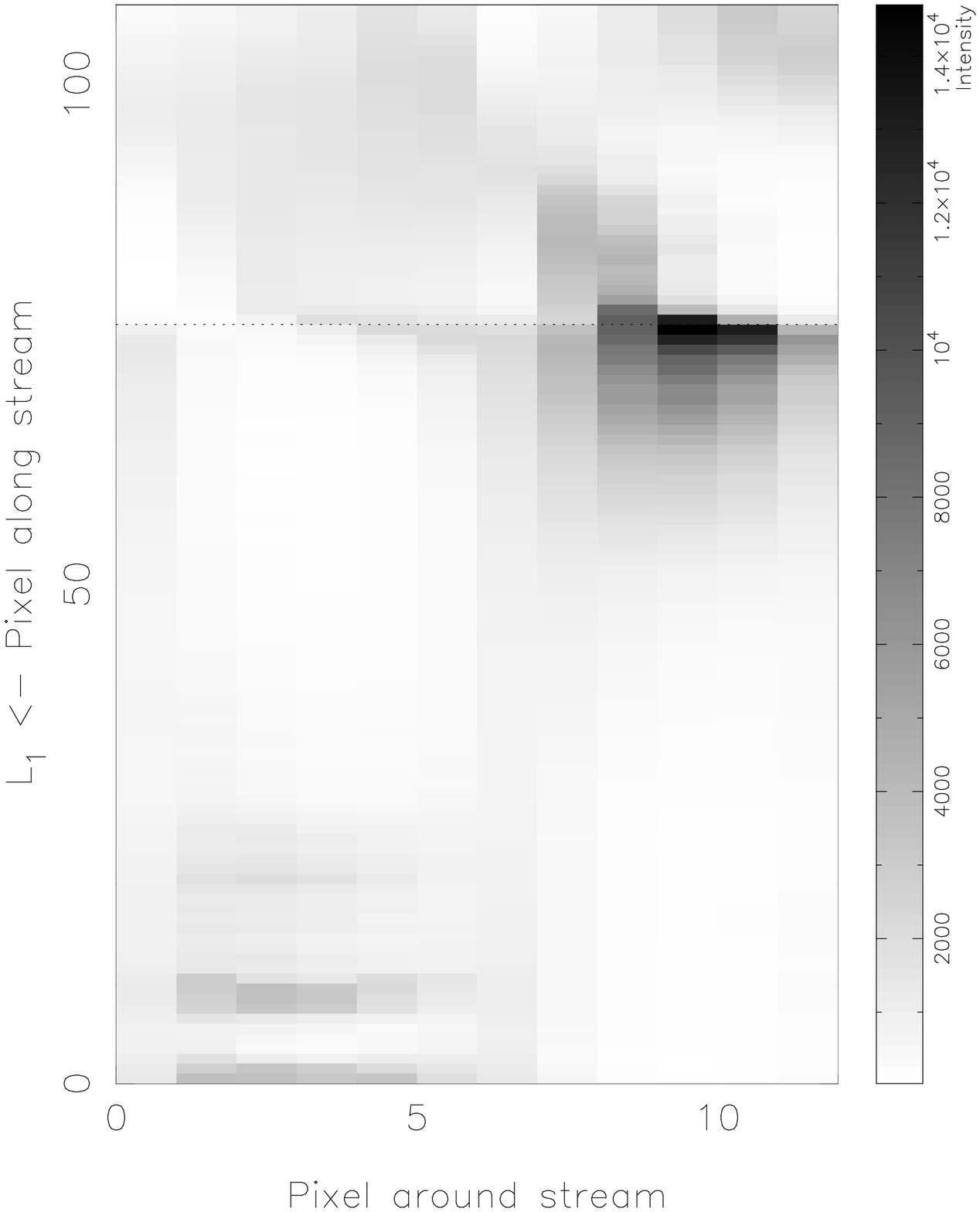,width=5.0cm,%
bbllx=47pt,bblly=34pt,bburx=567pt,bbury=692pt,clip=}
\end{minipage}
\hfill
\begin{minipage}{5.0cm}
\psfig{file=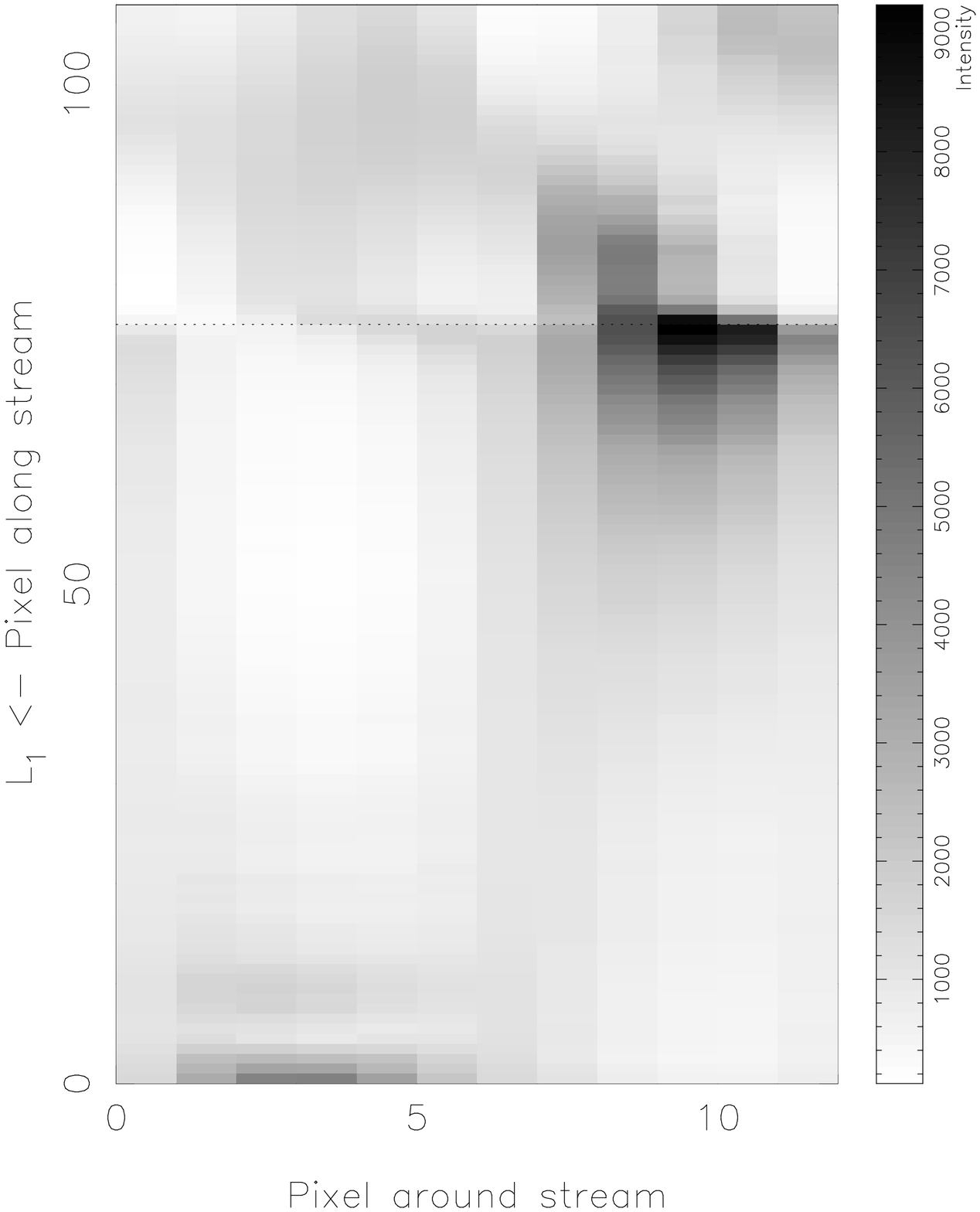,width=5.0cm,%
bbllx=47pt,bblly=34pt,bburx=567pt,bbury=692pt,clip=}
\end{minipage}
\hfill
\begin{minipage}{5.0cm}
\psfig{file=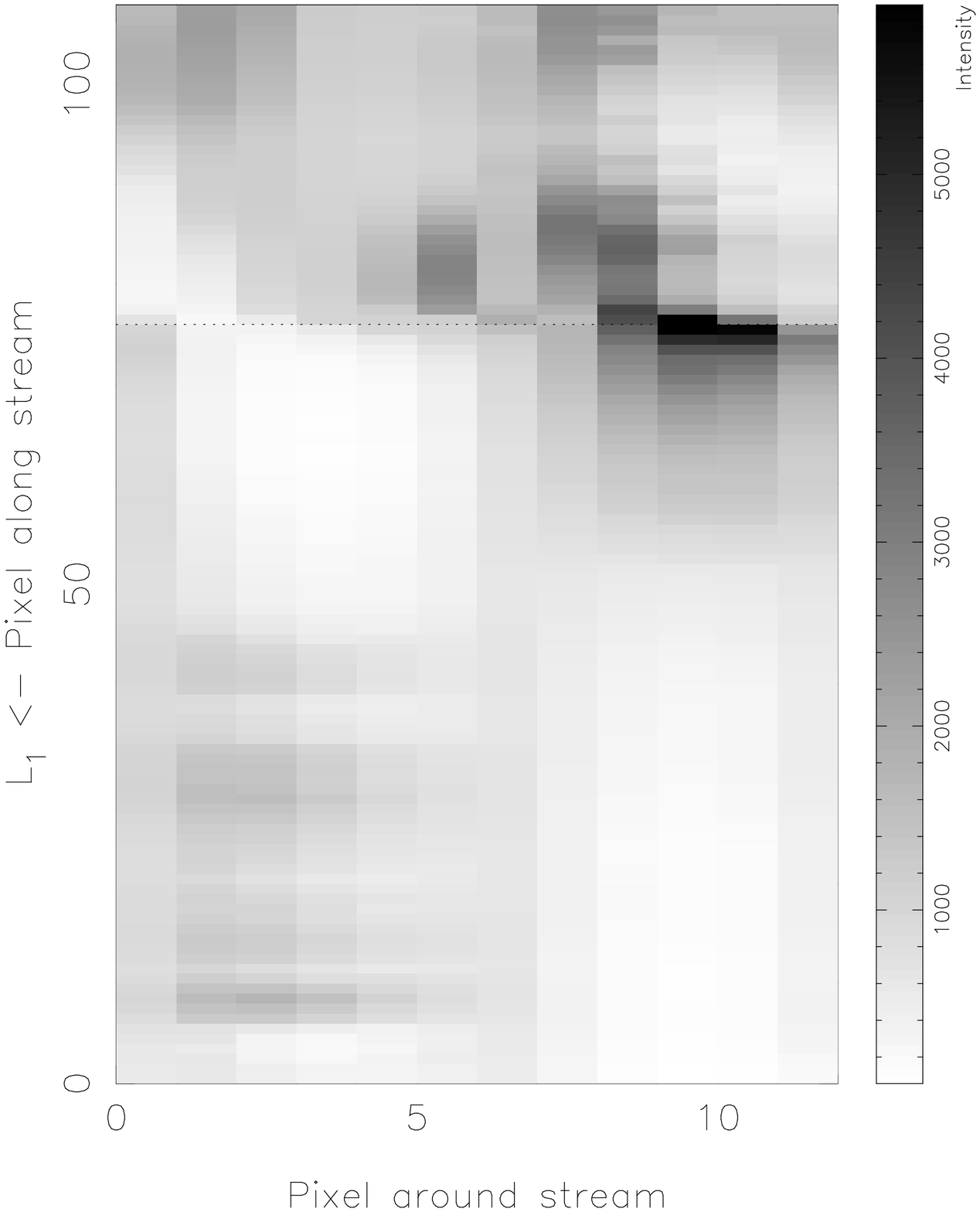,width=5.0cm,%
bbllx=47pt,bblly=34pt,bburx=567pt,bbury=692pt,clip=}
\end{minipage}

\caption{{\em Top row:} The light curves of H$\beta$, H$\gamma$ and \hel2\ as
analysed, i.e.\ corrected for the contribution of reprocessed light
from the secondary, together with the fits (solid line through
observed data) and the residuals (solid line around flux 0).
{\em Bottom row:} The recontructed intensity distributions on the surface of
the accretion stream as gray-sclae plots.The horizontal dotted line
at pixel 76 denotes the threading location.
\label{f:results}}
\end{figure*}
%***********************************************************************

%Hbeta eclipse************************************************************
\begin{figure*}
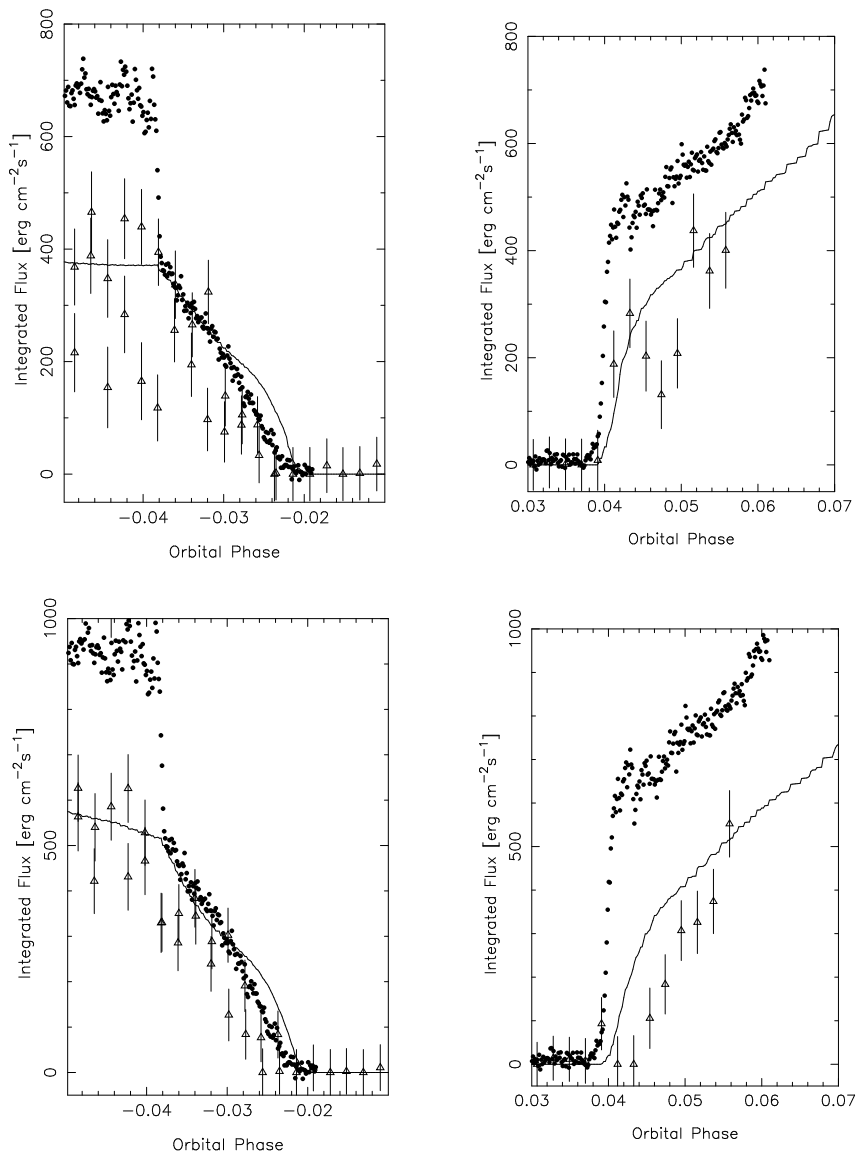

\noindent
\begin{minipage}{5cm}
\psfig{file=hb_fit_b_inf.ps,width=5cm}
\end{minipage}
\hspace{1cm}
\begin{minipage}{5cm}
\psfig{file=hb_fit_b_ef.ps,width=5cm}
\end{minipage}
%HeII eclipse************************************************************
\vspace{0.5cm}

\begin{minipage}{5cm}
\psfig{file=he_fit_b_inf.ps,width=5cm}
\end{minipage}
\hspace{1cm}
\begin{minipage}{5cm}
\psfig{file=he_fit_b_ef.ps,width=5cm}
\end{minipage}
\caption{Details of the light curves in H$\beta$ (top) and \hel2\
(bottom) at eclipse ingress and egress, respectively. Shown are the
integrated fluxes in the emission lines (triangles with error bars), 
the ASM-based fits to the light curves, the same as in Fig.~\ref{f:results}
(solid lines), and the broad-band $B$ light curves (filled circles).
\label{f:eclipse}}

\end{figure*}

\subsection{Application to HU~Aqr}\label{s:app_hu}

\subsubsection{The light curves}
The light curves of the three emission lines, corrected for the
contribution of reprocessed light from the secondary star, are shown
in Fig.~\ref{f:results} (top row). They are very similar in shape, showing the
total eclipse around phase zero, a secondary minimum at phase
$\sim$0.55 ($\equiv -0.45$), a primary maximum at phase $\sim$0.24, and a
secondary maximum at phase $\sim$0.74 ($\equiv -0.26$). The major difference
between the light curves of Hydrogen and Helium is the lower contrast
between the primary and secondary minima for the line of ionized
Helium. The sampling of the light curves is sparse in the phase
interval $0.05 - 0.08$, i.e. at phase of eclipse egress.

ASM was applied to these light curves and their overall shape was
reconstructed very well, except for some details of the eclipse
profiles (see below).  ASM tries to fit a given light curve until some
predefined value of $\chi^2 = \sum ((f - p)/\sigma)^2$ (where $f$ and
$p$ are the observed and predicted data, respectively and $\sigma$ is
the uncertaincy of the observed data points) is reached.  Since the
data points around eclipse are not very numerous, they do not affect
the general solution substantially.  The choice of the $\chi^2$'s
obviously introduces some kind of subjectivity. Our choice was driven
by the two opposing constraints to reach an as good as possible fit to
the light curves and not to introduce too many (most likely
artificial) structures into the maps.  Our finally chosen values of
$\chi^2 = 4$ for the Helium line and 7 for the Balmer lines appear
somewhat high. However, this choice was necessary due to the rather
large scatter of the individual data points
due to flickering and flaring in the lines and possibly small
deviations of the true stream geometry from the one we used.
For lower $\chi^2$, artefacts due to overfitting
were dominating the maps.

The fitted light curve at the given $\chi^2-$level is shown as solid
line in each panel of Fig.~\ref{f:results} with the same phase
sampling as the input data.  The solid line scattered around zero
intensity in each panel indicates the residuals of the data points
with respect to the fitting solution.  The residuals mainly show a
random scatter with an exception around eclipse phase, where the residuals
are systematically negative. Unfortunately, at egress, just where the
fit to the data would be especially important, the data are only
poorly sampled.

\subsubsection{The maps}

The reconstructions of all three emission lines clearly show a bright
spot in the threading region on that side of the stream which faces
the white dwarf. Otherwise, the intensity of the ballistic stream is
low, not unlike our test stream. The wave-like intensity pattern in
the lower left region of the maps, i.e.~near the $L_1$-point on the
non-illuminated side of the stream (circumference pixel $<7$, length
pixel $<30$) is interpreted as numerical artefact (see
Section~\ref{s:app_art}).

The fits show a migration of the maximum intensity from pixel '10' in the
threading region, where the ballistic stream is irradiated from the side
towards pixels '9' and later '8' in the magnetic part. In the twisted part
of the stream around the threading region these are the pixel facing the
white dwarf most perpendicularly, i.e.\ would receive most irradiation.
These pixel being so bright is therefore compatible with a direct
irradiation from the hot accretion spot where the ionizing source can be
located.

\subsubsection{Detail in the light curve}

Details of the light curves and the light-curve fits at eclipse
ingress and egress are shown together with simultaneous MCCP B-band
photometric data in Fig.~\ref{f:eclipse}. The MCCP data were scaled by
an appropriate factor so that the ASM-based fits and the MCCP data
agree with each other at the initial ingress phase (phase $\simeq
-0.038$). When comparing the two types of data, one should bear in
mind, that they consist of somewhat different radiation components:
The spectroscopic data contain only radiation from the emission lines
originating in the accretion stream.  This radiation is considerably
optically thick.  The MCCP data contain as well the recombination
continuum which is likely to be optically thin.  Outside the white
dwarf's eclipse, additional radiation from the undisturbed and
accretion-heated photosphere of the white dwarf and cyclotron
radiation from the accretion plasma is contained in the broadband data
recorded with the MCCP.  Hence, we cannot expect to see exactly the
same shapes of the emission line and broad-band photometric light
curves. However, these data with their high time resolution allow a
detailed comparison of the contact times and thus provide as with a
test if our overall geometry is chosen correctly.

First of all, the contact phases (however, not the egress slopes) of
the two MCCP-data and the ASM-fit agree with high precision.  For that
comparison one must not take notice of the initial steep steps at
eclipse ingress and egress in the MCCP-data, which are due to the
occultation of the white dwarf.  The precursor seen in the MCCP-data
immedidately before the steep rise at egress, 
$\phi = 0.037 - 0.039$,
is due to the non-heated photosphere 
of the white dwarf
emerging from eclipse. 
The spectroscopic data set and the model do not
contain radiation from the white dwarf surface, therefore, the egress
in the corresponding light curve is shallower and sets in at that
phase, when the spot reappears behind the companion star.
The contact time in the H$\beta$
light curve is consistent with the MCCP and ASM light curves, whereas
the He{\sc ii} light curve seems a bit late. We do not regard this as
a serious inconsistency. 
The reader shall be reminded, that each spectroscopic data point 
shown in Fig.~\ref{f:eclipse} represents a single spectrum of integration 
time 15 sec. The errors represent statistical errors only and do not account
for calibration uncertainties and the background subtraction. A further 
source of uncertainty is introduced by the continuum subtraction which
reveals slightly different results for the line fluxes 
if either a local or a global continuum is fitted and subtracted.
The points of contact can be compared reliably only between the model 
and the MCCP-data and these agree quite well which gives us confidence 
in our choice of geometry and system parameters.

Secondly, as mentioned earlier, the emission line light curves carry
less flux during stream ingress and egress than predicted by ASM. We
have no final explanation for this disagreement.  We cannot completely
exclude a measurement error since only a few spectra were obtained
particularly at eclipse egress.  Also, our assumption of complete
optical thickness might be a too strong requirement. Furthermore,
the stream almost certainly has a more complex geometry than being a
tube. SMH97 have shown that an accretion curtain exists and its
existence must have an effect on the light curves. Such structures are
probably best identified around eclipse. They do not necessarily alter
the contact times, the most crucial one being that at $\phi = -0.023$
which marks occultation of the outer envelope of the accretion
stream/curtain. But they influence the
overall brightness and the gradient
of the light curve.  If not a measurement error, we must conclude from
the described discrepancy that mapping experiments using eclipse light
curves alone will predict too high emission line fluxes outside
eclipse.

Thirdly, ASM predicts a bend of the eclipse light curves, most
pronounced at $\phi = -0.025 = 0.975$ and 0.045. Such bends are not
observed with the MCCP and it is difficult to determine, whether such
structure in the spectroscopic data set exists. With our assumed
stream geometry, the last structure visible at ingress phase before
being obscured by the companion star, is the unirradiated side of the
threading region. This side of the stream does not show a bright spot
like the irradiated side.  The sudden decrease of light predicted by
the ASM-fit after phase 0.975, is caused by the sudden eclipse of the,
however, still relatively bright region on the stream. In contrast,
the more gradual decline of the MCCP light curve might rather be
indicating emission from a more extended structure.  As suggested in
the preceding section, a structure like an accretion curtain, which
cannot be modelled with the present one-dimensional data set, might be
responsable for the discrepancy.

One should keep in mind, that the out-of-eclipse light curve is
significantly determined by the stream emissivity and is little
affected by the flickering and flaring or the low signal-to-noise
ratio. Even though the exact shape of the ingress and egress light
curves are not reproduced in detail and shape of this spot might not
be exactly reconstructed, we regard therefore the presence of a spot
at the threading region as highly likely.

\subsubsection{The influence of the correction function}

Finally, we would like to discuss the influence of the correction
function used for the contribution of reprocessed radiation from the
secondary star (see Fig.~7 in SM97). First of all, the fits at eclipse
phase are completely unaffected by the shape and scaling of the
correction function because the intensity of that radiation component
is zero at that particular phase interval. We investigated the
influence of the correction function by using differently scaled
correction functions of always the same shape (see
Section~\ref{s:data}).  We scaled the nominal function by factors 1.5
and 0.5, respectively.

When scaling with a factor of 1.5 there is only little intensity left
in the emission line light curves at phase 0.5. However, the effect on
the resulting maps are minor. Using the same $\chi^2$-levels as for
the fits in Fig.~\ref{s:app_hu} the same bright spots in the threading
region emerge where the spots are even more pronounced, i.e.~less
smeared.

If a scaling factor of only 0.5 is used (which is much less than the
actual uncertainty is), the maps show enhanced brightness on the
magnetic part of the stream on its outer side (pixel $2-6$). This is
easily explained: this part of the magnetic stream is best visible
around phase 0.5, when also the irradiated hemisphere of the secondary
star is best visible. Since our code does not account for the surface
of the secondary star the stream apparently emits excess radiation at
phase 0.5. This excess radiation is projected onto the outer magnetic
stream, a physically unlikely location.

\section{Discussion}\label{s:disc}

\subsection{The bright spot on the accretion stream}

Our ASM-code predicts a bright spot of line radiation in the threading
region, where the stream is redirected and couples to magnetic field
lines.  This implies that X-ray/EUV irradiation alone is seemingly not
the major source of excitation and ionization, but the combination of
irradiation plus dissipative heating seems to be able to do the job.
To our knowledge, this is, together with the work by Hakala (1995) and
Harrop-Allin et al.~(1998), the first direct proof of the prediction
made further back in past by Liebert \& Stockman (1985), that
dissipation in the threading region occurs and has observational
consequences. However, their prediction of X-ray emission from the
threading region is still missing and can be proven only by
XMM-observations of eclipsing polars.

A similar bright spot as in our brightness maps was also found by
Hakala (1995) in observations of \hu\ in a low accretion state. Apart
from that, Hakala reports a pronounced brightness increase towards the
white dwarf. In their analysis of the 1993 high state $UBVR$ light
curves, Harrop-Allin et al.~(1998) found the ballistic stream to be
bright near the $L_1$-point and faint in the threading region. Just
after the threading region towards the white dwarf, they observed a
sudden brightness increase in all four colours.  In contrast, their
1999 mapping analysis (Harrop-Allin et al.~1999b) of the same data
shows an overall faint ballistic stream except for a bright spot near
the threading region. As their one-footpoint stream leaves the orbital
plane it fades and re-brightens again as it approaches the white
dwarf. The reason for their picture changing so dramatically (as far
as the ballistic stream is concerned) is a change of their global
optimization scheme.  Since their more recent analysis shows results
closer to ours, we regard this as the more likely solution.  Our and
their analysis of different data sets taken only a few days apart show
commonly the stream to be bright in the threading region. However,
Harrop-Allin et al.\ used a restricted data set centred on the white
dwarfs eclipse.  This, together with the simplifying assumption of
constant cyclotron and photospheric brightness out of eclipse, are the
likely reasons for the differences in the resulting maps.

\subsection{The optical depth in the stream}

In our approach the stream is assumed to be completely optically thick 
in the lines. This approach is supported by the fact that the observed 
emission line light curves show two distinct maxima at those phases
when the largest surface of the stream is offered to the observer.

A comment concerning the recent work by Harrop-Allin et al.~(1999b),
which appeared after submission of the first version of our paper:
They used broad-band data observed quasi-simultaneously with our
spectroscopic data and mapped the brightness on a one-dimensional
accretion stream. They also assumed optically thick radiation,
even in the continuum.  However, in their maps they assign only one
brightness value to a certain pixel along the accretion stream. Both
ingress and egress were used, where either the irradiated
or the unirradiated side of the stream is seen. As it is clear from
our emission line light curves and from our maps, the stream has
different brightness on both sides. In this respect the approach by
Harrop-Allin et al.~(1999b) is insufficient.

Along the same line, ASM has also an advantage over Doppler
tomography: in Doppler tomography one implicitely assumes that the
emission is optically thin. Therefore, no information about different
emissivities from the irradiated and non-irradiated sides can be
obtained from the Doppler maps.

Furthermore, our ASM approach has the advantage of using the maximum
information in the form of the full orbital light curve. Using
emission line data, we exclude explicitely emission from the white
dwarf and the accretion spot. This offers a much more undisturbed and
cleaner view onto the accretion stream. Our presented analyses
suffers, however, from the few photons available around
eclipse phase, where -- at least in principle -- crucial tests of the
stream geometry and brightness distribution were possible.

\subsection{The luminosity balance of the stream}

As mentioned earlier, Hakala's and Harrop-Allin et al.'s maps show
dominating brightness in the magnetic part of the stream close to the
white dwarf. Fig.~\ref{f:eclipse} shows, that the $B$ band light
curve displays a more gradual decline during ingress and egress than
predicted by ASM. These observed profiles may be caused by a 
brighter magnetic stream than our mapping experiment predicts
but may also be due to a bright accretion curtain or a similar more
complicated geometric structure.

However, we searched for independent information about the
luminosity balance between the ballistic and the magnetic streams by
analysing the observed emission line profiles directly.  We used two
approaches: in the first we calculated the integrated brightness in
selected parts of the Doppler maps; in the second we decomposed the
line profiles at appropriate phases into subcomponents.  Both
approaches rely to a large extent on the assumption that the observed
structures in the Doppler maps and in the spectral lines can be
assigned to physically different locations in the binary system. 

In the Doppler maps three distinct structures can be associated to the
irradiated secondary star, the ballistic stream and the magnetic
stream (SMH97, Figs.~11, 12). The Doppler map of \hel2\ suggests that
about one third ($\sim$30\%) of the total stream emission originates from its
magnetic part.

Investigating the emission lines is somewhat more complicated:
The parallel projection of a Doppler map at a given angle gives a
spectrum (intensity vs.~velocity, hence wavelength) at a certain
binary phase, as observed in the line profile.  The structure of the
Doppler maps presented by SMH97 show clearly that no projection angle
exists where the different structures can be separated uniquely into
spectral features, i.e.~without contamination by a another emission
component. However, the separation between the ballistic and the
magnetic parts of the stream seems to be most easily possible at
phases 0.25 and 0.75.

At these phases, the ballistic stream is perpendicular to the line of
sight to the observer. The velocity component we see as the radial
velocity is $v_y$ in the Doppler maps and which is relatively sharply
defined. It can be read off from the Doppler map as $v_y \simeq
+190$\,\kmps. The magnetic stream occupies the region of the
$(v_x,v_y)-$Doppler map with negative $v_y-$velocities, because at the
given phases the radial velocities along the magnetic stream are
antiparallel to those of the ballistic part. Hence, in the line
profile at the given phases it occurs as a broad feature with opposite
sign of its bulk velocity (opposite sign with respect to the ballistic
stream).  Even at these optimum phases the high-velocity tail of the
ballistic stream overlaps with emission from the magnetic stream (in
the parallel projection of the Doppler map).  Therefore, a
decomposition using Gaussian curves will result in an overestimate of
the contribution from the magnetic stream.  A Gaussian deconvolution
of the spectral line profiles of \hel2\ at the given phases reveals
less than $\sim$50\% emission from a broad base component. The
percentage is somewhat lower for H$\beta$, namely $\sim$45\%.

As expected, the fraction of flux in the broad base component (i.e.\
originating in the magnetic stream) is higher than the value derived
from the measurement made in the Doppler map, due to the superposition
of different radiation components.  We conclude, that observationally
the contribution of light from the magnetic stream to the total
brightness of the stream is certainly below 50\%, but more likely of
the order of 30\%. The integrated brightness from the magnetic stream
of our maps (Fig.~\ref{f:results}) is $\sim$40\% in \hel2\ and
$\sim$30\% in the Balmer lines.  We regard these numbers as support
for our approach.

\subsection{Outlook}

Clearly, for a better understanding of the brightness distribution
along the different parts of the accretion stream and the accretion
curtain one needs to take into account not only the light curve,
but also the line profiles, i.e.\ combine the two methods {\em Doppler
Tomography} and {\em Accretion Stream Mapping}.  We plan to establish
an {\it Eclipse Doppler tomography} as a future step of analysis.  The
necessary input data with high time and spectral resolution can be
obtained utilizing 8m-class telescopes.

\section*{acknowledgments}
We thank the referee for helpful comments. SV thanks the South African
NRF (formerly FRD) for funding a postdoc fellowship and the AIP for
kind hospitality in October 1997 and April 1999. This work was
supported by DLR grant 50 OR 9706 8.

\end{document}